\documentclass[12pt]{article}
\usepackage{latexsym,graphicx,multirow}
\usepackage{float}
\usepackage{amssymb}
\usepackage{amsmath}
\usepackage{amscd}
\usepackage{amsthm}
\usepackage[left=2cm,top=2.5cm,right=2.5cm,bottom=1.5cm]{geometry}
\usepackage{hyperref}
\usepackage{epstopdf}
\usepackage{cite}
\usepackage{caption}
\usepackage{subcaption}

\begin{document}
	
\begin{center}
\large{\bf{  Bianchi type-V Transitioning  model in Brans-Dicke theory with Observational Constraint}} \\
		\vspace{10mm}
\normalsize{Vinod Kumar Bhardwaj$^1$, Archana Dixit$^2$, Anirudh Pradhan$^3$ }\\
		\vspace{5mm}
\normalsize{$^{1,2}$Department of Mathematics, Institute of Applied Sciences and Humanities, GLA University, Mathura-281 406, Uttar Pradesh, India}\\
\vspace{5mm}
\normalsize{$^{3}$Centre for Cosmology, Astrophysics and Space Science (CCASS), GLA University, Mathura-281 406, Uttar Pradesh, India}\\	
		\vspace{2mm}
$^1$E-mail:dr.vinodbhardwaj@gmail.com\\
		\vspace{2mm}
$^2$E-mail:archana.dixit@gla.ac.in\\
		\vspace{2mm}
$^3$E-mail:pradhan.anirudh@gmail.com \\
		\vspace{2mm}

\end{center}
	
\begin{abstract}

In this paper, we have examined the viability of the Bianchi type-V universe in Brans-Dicke (BD) theory of gravitation. We have discussed the interacting and non-interacting scenarios between dark matter (DM) and dark energy (DE) of the derived universe within the framework of BD theory. 
CCA technique has been applied to constrain the model parameters using 46 values of observational Hubble data (OHD), Pantheon data (the latest compilation of SNIa with 40 binned in the redshift range $0.014 \leq z \leq 1.62)$ and their combined datasets.  We establish an exact solution of the field equations to derive the dynamics of the derived universe and the obtained results are found to agree with the observations
We also noted a distinctive change in the sign of the deceleration parameter from positive to negative, as well as the presence of a transition red-shift exists. Using various observational data points, the evolution trajectories for $(r~- s)$ diagnostic planes are shown to understand the geometrical behavior of the Bianchi-V model. Some physical properties of the universe are also discussed. It's also worth noting that the conclusions of the cosmological parameter are consistent with modern observational data.

\end{abstract}
	
\smallskip 
{\bf Keywords} : LRS Bianchi-V, Brans-Dicke theory, $ \Lambda$CDM Model, Statefinders.
	
PACS: 98.80.-k \\
	
\section{Introduction}

According to recent studies of type Ia supernovae \cite{ref1,ref2}  our current universe is undergoing an accelerated phase of expansion proceeded 
by a period of deceleration. To examine the current phase of acceleration, a new form of substance with negative pressure, called dark energy, has 
been suggested. In literature several models have been examined, to predict cosmic acceleration by assuming dark energy with repulsive gravity as 
the major content of the cosmos. According to the Planck observations, Universe’s total mass-energy budget is composed of 68.3 \% DE, 26.8 \% 
dark matter and 4.9 \% ordinary matter. This exotic form of energy is known as dark energy which is to be less effective in early times but 
dominates at the present epoch. The attracting force of DM is involved in the creation of structure and clustering of galaxies.\\

Although several observational experiments support DE as a general clarification for the expansion mystery, but there are many questions which 
remain unanswered. The cosmological constant ($\Lambda$) has a repulsive nature, it is a usual candidate for DE. It posseses a low magnitutde in 
comparison with prediction of particle physics. Apart from dark energy models, a variety of gravity theories such as f (R) gravity 
\cite{ref3,ref4},$f(T)$ gravity  \cite{ref5}, and $f(G)$ gravity \cite{ref6,ref7}, $f(Q)$ gravity \cite{ref7a,7b}, $f(Q, T)$ \cite{ref7c}, Einstein-Gauss-Bonnet theory \cite{ref7d,ref7e,ref7f,ref7g,ref7h}  were discussed. In the same direction, $f(R,T)$ theory is the 
another perspective of modified version of theory of gravity, which has been  proposed by Harko et al.\cite{ref8}. Similaly,  many theories such 
as “scalartensor theories, scalar field theories, bimetric theories and vector-tensor theories \cite{ref9,ref10}, Weyl theory \cite{ref11}, 
Lyra theory \cite{ref12} Ylmaz theory \cite{ref13} and Brans-Dicke theory \cite{ref14,ref15}” are some examples of these alternative gravitation 
theories \cite{ref16}. For the previous coupled of decades, cosmologists have been interested in alternative theories of gravity, particularly "scalar-tensor" theories. Among all these alternative theories, the ``Brans-Dicke theory (BDT)” is the one of the most successful alternate theory. Alonso et al. \cite{ref16a} have shown that the scalar tensor theory of Brans-Dicke type provides a relativistic generalization of Newtonian gravity in 2+1 dimensions. The theory is metric and test particles follow the space-time geodesics. \\

BD theory, on the other hand, is often considered as the prototype of a wide category of gravitational theories known as scalar-tensor theories. These theories are derived through a non-minimal coupling between the curvature scalar and a scalar field. These hypotheses have gotten a lot of interest in the cosmology community because they can accurately represent the early expansion of the universe \cite{ref16b}. The BD hypothesis predicts an extended period of inflation to address the universe's graceful exit dilemma, as outlined by Mathiazhagan and Johri \cite{ref16c}, La and Steinhardt \cite{ref16d}. Furthermore, with suitable parameter values, the theory can successfully construct late-time accelerated expansion without the contribution of DE (\cite{ref16e})
Several authors have been explored BD cosmology in different prospectives. The work of Singh and Rai  \cite{ref17} provides a detailed 
discussion of Brans-Dick cosmological models. Remarkable studies of BD theory  with the application of  a non-vanishing cosmological constant 
can be seen in \cite{ref18,ref19,ref20,ref21,ref22,ref23,ref24}. In BD theory, authors \cite{ref25,ref25a,ref25b} studied a model of accelerating universe with 
changeable deceleration parameter. In 1984 Brans-Dike cosmological model was investigated by taking $\Lambda$. The same idea was further extended 
inhomogeneous and anisotropic Bianchi type I spacetime \cite{ref26,ref27}.
Akarsu, et al. \cite{ref27a} studied the anisotropic $\Lambda$CDM model with Brans–Dicke gravity. Recently, exact solutions of accelerating cosmological models in modified Brans-Dicke theory are proposed \cite{ref27b,ref27c}. \\

In cosmology, Bianchi-type space-times play a very important role in comprehending and describing the early phases of the universe's evolution. 
The study of Bianchi types II, VIII, and IX  universe is particularly relevant since they correlate to well-known solutions such as the FRW 
universe with positive curvature, the de Sitter universe, the Taub-NUT solutions, and so on. The Bianchi type-V frame of reference is very 
important for developing the models for measuring the expansion of the Universe in its early stages \cite{ref28,ref29}.  Several authors \cite{ref30,ref31,ref32,ref32a,ref32b,ref32c} have been interested in ``Bianchi type-V space-time type over the last decade. The dynamical behavior of the Bianchi-V model is significantly more general than the simple FLRW model. The study of Bianchi type-V cosmological models creates more interest 
as these models contain isotropic special cases and permit arbitrarily small anisotropic levels at certain stages \cite{ref33}. However, we know that, 
in recent years, Bianchi cosmological models have become extremely important in observational cosmology, as evidenced by the WMAP observational data. 
It has validated an addition to the standard ``$\Lambda CDM$" model that resembles the Bianchi morphology. The WMAP data analysis supports the 
fact that the universe has a preferred direction and it should reach a slightly anisotropic geometry. Therefore, in view of the background anisotropy, 
the models with anisotropic backgrounds are more suitable to describe early stages of the universe \cite{ref34,ref35}.\\
 
Cosmology is now one of the fastest-developing branches of science, particularly in terms of observations. In the last decade, cosmological 
data from various sources has been rapidly updated, not only in terms of quantity but also in terms of quality \cite{ref36}. The latest ``CMB anisotropies" results by ``Planck Collaboration" provide the precise observations on temperature and polarization of the photons from the last scattering surface near the redshift about $z = 1090$ \cite{ref37}. Another observable, which is worthy of mentioning, is the ``redshift-space distortion (RSD)". It has been measured more and more precisely in the past few years. This observable is a key signature to disclose the large-scale structure. In the present study, we proposed transitioning model of the universe Bianchi-V space-time in the framework of  BD theory. To achieve the explicit solution for the model,  we  assumed the scalar field 
$\phi = \phi_{0}(t) [(k_{1}e^{n\beta t}-1)^{1/n}]^{\eta} $; where $\phi_{0}$, $k_{1}$, $n$, $\beta$ and $\eta$ are constants.  In the present study, the recent OHD and Pantheon data sets are used to decide the model parameters`` $n$ and $\beta$ \cite{ref38}. We investigate a transitioning model of the universe for interacting and non-interacting scenarios within the framework of Brans-Dicke theory.\\
 
In our derived model, the best fit values of parameters are determined by using
$\chi^{2}$ statistic. Two observational data sets, i) A data set of 46 OHD and ii) Pantheon data (the latest compilation of SNIa with 40 binned in the redshift range $0.014 \leq z \leq 1.62)$ are considered to discuss the dynamics 
of the derived universe. This paper is structured as follows: In section 2, we introduce the basic field equation of the Bianchi- V universe. In Section 3, 
we discussed the observational constraints of the model parameter.  In Section 4, we have discussed the non-interacting and interacting models. 
The physical behavior of the model is analyzed in Section 5. In Section 6, we apply the Statefinder diagnostic. Conclusions are mentioned 
in section 7.

\section{Field Equations of Bianchi type-V model}

The action for Brans-Dicke theory is given by

\begin{equation} \label{1}
S=\frac{1}{16 \pi} \int {\sqrt{-g}  \left(\phi R-\omega \frac{\phi_{,i}\phi^{,i}}{\phi}  \right) d^{4}x }+\frac{1}{16 \pi} \int {\sqrt{-g} L_{m}  d^{4}x},
\end{equation}
where $ L_{m} $ stands for Lagrangian matter field, $ \phi $ denotes the Brans-Dicke scalar field and $ \omega $ is the Brans-Dicke coupling constant.\\

The field equations in Brans-Dicke theory \cite{ref39,ref40} by assuming dark matter (DM) and dark energy (DE) are read as:

\begin{equation}\label{2}
R_{ij}-\frac{1}{2} R g_{ij}=-\frac{8 \pi}{\phi} \left(T^{m}_{ij}+T^{de}_{ij}\right)-\frac{\omega}{\phi^2}\left(\phi_{i}\phi_{j}-\frac{1}{2} g_{ij} 
\phi_{k} \phi^{k}\right)-\frac{1}{\phi} \left(\phi_{ij}-g_{ij} \Box\phi\right),
\end{equation}
and 
\begin{equation}\label{3}
\Box\phi = \frac{8 \pi }{(3+2 \omega)} \left(T^{m}+T^{de}\right),
\end{equation}
where $ T^{m}_{ij}= dia\left[-1, \omega_{m}, \omega_{m}, \omega_{m}\right]\rho_{m} $ and  
$ T^{de}_{ij} = dia\left[-1, \omega_{de}, \omega_{de}, \omega_{de}\right]\rho_{de} $ 
are respectively the energy momentum tensors for DM and DE.\\

The energy conservation equation is read as
\begin{equation}\label{4}
\left(T^{m}_{ij}+T^{de}_{ij}\right)_{;j}=0,
\end{equation}
which leads to
\begin{equation}\label{5}
\dot{\rho_{de}}+3H(1+\rho_{m}+\omega_{de}) \rho_{de}+\dot{\rho_{m}} =0
\end{equation}

We consider the Bianchi type-V metric of the form
\begin{equation}\label{6}
ds^2 = dt^2 - A(t)^2 dx^2 - e^{-2\alpha x} \left[B(t)^2 dy^2+C(t)^2 dz^2 \right],
\end{equation}
where $\alpha$ is a constant and the functions $A(t)$, $B(t)$ and $C(t)$ are the three anisotropic directions of expansion in normal three 
dimensional space. Those three functions are equal in FRW models due to the radial symmetry and so we have only one function 
$a(t)$ there. The average scale factor $a$, the spatial volume $V$ and the average Hubble's parameter 
$H$ are defined as $a=(ABC)^{1/3}$, $V= a^{3} = ABC$, and $H= \frac{1}{3}\left(\frac{{\dot{A}}}{A} +\frac{{\dot{B}}}{B} + \frac{{\dot{C}}}{C}\right)$ 
respectively.\\

In a comoving coordinate system, the field equations given by Eqs. (\ref{2}) and (\ref{3}), for Bianchi type-V spacetime mentioned in Eq. (\ref{6}) with 
energy-momentum tensors defined previously, are read as

\begin{equation}\label{7}
\frac{\ddot{B}}{B}+\frac{\ddot{C}}{C}+\frac{\dot{B} \dot{C}}{BC}-\frac{\alpha^2}{A^{2}}+\frac{\omega}{2} \frac{\dot{\phi}^2}{\phi^2}+
\frac{\dot{\phi}}{\phi} \left(\frac{\dot{B}}{B}+\frac{\dot{C}}{C}\right)+\frac{\ddot{\phi}}{\phi}=- \phi^{-1} p_{de}
\end{equation}\label{8}
\begin{equation}
\frac{\ddot{C}}{C}+\frac{\ddot{A}}{A}+\frac{\dot{C} \dot{A}}{CA}-\frac{\alpha^2}{A^{2}}+\frac{\omega}{2} \frac{\dot{\phi}^2}{\phi^2}+
\frac{\dot{\phi}}{\phi} \left(\frac{\dot{C}}{C}+\frac{\dot{A}}{A}\right)+\frac{\ddot{\phi}}{\phi}=- \phi^{-1} p_{de}
\end{equation}
\begin{equation}\label{9}
\frac{\ddot{A}}{A}+\frac{\ddot{B}}{B}+\frac{\dot{A} \dot{B}}{AB}-\frac{\alpha^2}{A^{2}}+\frac{\omega}{2} \frac{\dot{\phi}^2}{\phi^2}+
\frac{\dot{\phi}}{\phi} \left(\frac{\dot{A}}{A}+\frac{\dot{B}}{B}\right)+\frac{\ddot{\phi}}{\phi}=- \phi^{-1} p_{de}
\end{equation}
\begin{equation}\label{10}
\frac{\dot{A} \dot{B}}{AB}+\frac{\dot{B}\dot{C}}{BC}+\frac{\dot{C} \dot{A}}{CA}-\frac{3\alpha^2}{A^{2}}-\frac{\omega}{2} 
\frac{\dot{\phi}^2}{\phi^2}+\frac{\dot{\phi}}{\phi} \left(\frac{\dot{A}}{A}+\frac{\dot{B}}{B}+\frac{\dot{C}}{C}\right)= \phi^{-1} 
\left(\rho_{m}+\rho_{de}\right)
\end{equation}
\begin{equation}\label{11}
2\frac{\dot{A}}{A}-\frac{\dot{B}}{B}-\frac{\dot{C}}{C}=0
\end{equation}
\begin{equation}\label{12}
\frac{\ddot{\phi}}{\phi}+\frac{\dot{\phi}}{\phi} \left(\frac{\dot{A}}{A}+\frac{\dot{B}}{B}+\frac{\dot{C}}{C}\right)=
\frac{\rho_{de}(1-3\omega_{de})+\rho_{m}}{(3+2\omega)}
\end{equation}
From Eqs. (\ref{7})- (\ref{9}), we have 
\begin{equation}\label{13}
\frac{\ddot{A}}{A}+\frac{\dot{A}\dot{C}}{AC}-\left(\frac{\ddot{B}}{B}+\frac{\dot{B}\dot{C}}{BC}\right)+\frac{\dot{\phi}}{\phi}
\left( \frac{\dot{A}}{A}-\frac{\dot{B}}{B}\right) =0
\end{equation}

\begin{equation}\label{14}
\frac{\ddot{B}}{B}-\frac{\ddot{C}}{C}+\frac{\dot{A}\dot{B}}{AB}-\frac{\dot{A}\dot{C}}{AC}+\frac{\dot{\phi}}{\phi}
\left( \frac{\dot{B}}{B}-\frac{\dot{C}}{C}\right) =0
\end{equation}

\begin{equation}\label{15}
\frac{\ddot{C}}{C}-\frac{\ddot{A}}{A}+\frac{\dot{B}\dot{C}}{BC}-\frac{\dot{A}\dot{B}}{AB}+\frac{\dot{\phi}}{\phi}
\left( \frac{\dot{C}}{C}-\frac{\dot{A}}{A}\right) =0
\end{equation}

Integrating Eq. (\ref{11}) and omitting constant of integration, we get
\begin{eqnarray}\label{16}
&BC = A^{2} \nonumber\\
&\Rightarrow  C = A D, \  \& \ B= A/D,
\end{eqnarray}
here $ D=D(t)$ denotes the measures of anisotropy in the Bianchi-V model.\\

From  Eqs. (\ref{14}) and (\ref{16}), we obtain
\begin{equation}\label{17}
\left( \frac{\ddot{D}}{D}-\frac{\dot{D}^{2}}{D^{2}}\right)+\frac{\dot{D}}{D}\left( 3\frac{\dot{A}}{A}+\frac{\phi}{\phi} \right)=0,
\end{equation}
which on integration reduces to

\begin{equation}\label{18}
D=exp\left( \int c_{1} A^{-3} \phi^{-1} dt\right) 
\end{equation}
where $ c_{1} $ is the integration constant.\\

Usng Eq. (\ref{16}), we can determine the scale factor as $ a^{3}= A^{3} \Rightarrow a=A $.\\

Thus, we get the following equations finally,
\begin{equation}\label{19}
2\frac{\ddot{a}}{a}+\frac{\dot{a}^2}{a^2}-\frac{\alpha^{2}}{a^2}+\frac{c_{1}^2}{a^{6} \phi^{2}}+\frac{\omega}{2} 
\frac{\dot{\phi}^2}{\phi^2}+2 \frac{\dot{\phi}}{\phi} \frac{\dot{a}}{a}+\frac{\ddot{\phi}}{\phi}=- \phi^{-1} p_{de},
\end{equation} 
\begin{equation}\label{20}
3 \frac{\dot{a}^2}{a^2}-3\frac{\alpha^{2}}{a^2}-\frac{c_{1}^2}{a^{6} \phi^{2}}-\frac{\omega}{2} \frac{\dot{\phi}^2}{\phi^2}+
3 \frac{\dot{\phi}}{\phi} \frac{\dot{a}}{a}=\phi^{-1} \left(\rho_{m}+\rho_{de}\right),
\end{equation} 
\begin{equation}\label{21}
\frac{\ddot{\phi}}{\phi}+3\frac{\dot{\phi}}{\phi} \frac{\dot{a}}{a}=\frac{\rho_{de}(1-3\omega_{de})+\rho_{m}}{(3+2\omega)}.
\end{equation}
On solving these equation we get dyanamic features of the proposed model.\\

Authors\cite{ref41,ref42} have considered the funtional relation of Brans-Dicke scalar field $ \phi $ and scale factor $ a $ as
\begin{equation}\label{22}
\phi=\phi_{0} a^{\eta},
\end{equation}
where  $ \eta$  and $\phi_{0}$ are constants. Choice of $ \phi=\phi_{0} a^{\eta} $ leads to be consistent  with results \cite{ref43}. \\

The observations of type Ia supernovae \cite{ref44,ref45,ref46}, ``CMB anisotropies'' \cite{ref47}, and the 
``Planck Collaborations" \cite{ref48} have verified that the universe is currently expanding speedly, which was decelerating in past. As a result, 
the universe must have a ``signature flipping"  from past deceleration to current acceleration \cite{ref49,ref50}. To determine the 
explicit solutions of transitioning universe, we have considered a special parameterization of the Hubble parameter \cite{ref50a,ref50b,ref50c}  
$H=\frac{\dot a}{a}=\beta (1+a^{-n})$, where $\beta>0$ , $n>1$ are constants.
On integration of this parametrization, we get an explicit form of the scale factor as $a(t) = \left(k_{1}e^{n\beta t} -1\right)^{1/n}$; where $k_{1} > 0 $ is the constant of integration \cite{ref38}. The present explicit form of scale factor is an exponential function containing two model parameters $n$ and $\beta$ which describe the dynamics of the universe. As $t \to 0$, we can have $a(0)= (k_{1} - 1)^{1/n}$, which provides a non-zero initial value of scale factor for $k_{1} \ne 1 $ (or a cold initiation of universe with finite volume).\\
Now, the Brans–Dicke scalar field $ \phi $ reads as
\begin{equation}\label{23}
\phi=\phi_{0} \left[\left(k_{1} e^{n \beta t} -1\right)^{\frac{1}{n}} \right]^{\eta}
\end{equation}
The deceleration parameter (DP) $q$ is obtained as

\begin{equation}\label{24}
q = -\frac{a\ddot{a}}{\dot{a}^2}=-\left( 1+\frac{\dot{H}}{H^2}\right) = -1 + \frac{n}{k_{1}e^{n \beta t}}
\end{equation}
Eq. (24) shows that the DP is time-dependent, which can take both positive and negative values representing early decelerating phase and later accelerating phase. From Eq. (24) we can see that $q = -1 + \frac{n}{k_{1}}$ as $t \to 0$, which is constant and positive for $n>1$ and $k_{1} < n$ and for $k_{1} > n$, DP possesses negative value. This indicates that DP has a signature flipping nature from positive to negative era with the evaluation.\\
By using the relation $(1+z) = \frac{a_{0}}{a}$, we express the cosmological parameters in terms of redshift. Thus, by applying the above transformation, $t$ in terms of $z$ can be found as $\frac{1}{n\beta}log\left[\frac{1+(1+z)^{n}}{k_{1}(1+z)^{n}}\right]$.  To describe the dynamics of the Universe, the Hubble parameter $H$ in terms of redshift can be read as 

\begin{equation}\label{25}
	H(z)=\beta [(1+z)^{n}+1].
\end{equation}

or
\begin{equation}\label{26}
	H(z)=\frac{H_{0}}{2}(1+(1+z)^{n}).
\end{equation}
In order to describe the dynamics properties of the model and the bahaviour of physical parameters, we discuss two observational data sets 
in following section.

\section{Observational constraints of the model }
\subsection{Observational Hubble data set (OHD)}

In this section, we have estimate the model parameters $ n $ and $H_0$ using recent 46 points of the $ H(z)$ data set (OHD) in the red-shift range $ 0 \leq z \leq 2.36 $ and their corresponding standard deviation $\sigma_{i} $. These 46 OHD points are summarized in the Table-I \cite{ref40}-\cite{ref59}. The best fit value of model parameters $n$ and $H_{0}$ obtained using  $\chi^2$ statistic, which is equivalent to the maximum likelihood analysis  given as 

\begin{equation}\label{27}
\chi^{2}_{OHD}\left( n, H_{0}\right)=\sum_{i=1}^{46} {\frac{\left(H_{th}(n,H_{0}, z_{i})-H_{ob}(z_i)\right)^2}{\sigma(i)^2}},
\end{equation}
where $ `H_{th}'$ and $ `H_{obs}' $ denote respectively the theoretical and observed value of Hubble parameter $ H $. By analysis, we found 
the values of parameters $n=1.457\pm0.042 $ and$ H_{0} = 68.46\pm1.97$ $km/s/Mpc$, which are comparable 
with the current Hubble parameter from Planck 2014 results \cite{ref61}. 

\begin{table}[H]
	\caption{\small The OHD data set of  Hubble parameter $H(z)$}
	\begin{center}
	\begin{tabular}{|c|c|c|c|c|c|c|c|c|c|}
	\hline
	\tiny	$S.No$  &	\tiny  $Z$ & \tiny $H (Obs)$ & \tiny $\sigma_{i}$ & \tiny References & \tiny $S.No$  &	\tiny  $Z$ & \tiny $H (Obs)$ & \tiny $\sigma_{i}$ & \tiny References \\
	\hline
	\tiny	1	& \tiny 0	  &\tiny 67.77 & \tiny 1.30 & \tiny\cite{ref51} & \tiny	24	& \tiny 0.4783	 &\tiny 80.9  & \tiny9 & \tiny \cite{ref48}   \\
		
			\tiny	2	& \tiny 0.07  &\tiny 69    & \tiny 19.6 & \tiny \cite{ref52} & \tiny	25	& \tiny 0.48	 &\tiny 97 & \tiny60 & \tiny \cite{ref43}  \\

			\tiny	3	& \tiny 0.09	 &\tiny 69  & \tiny 12 & \tiny \cite{ref53}   & 	\tiny	26	& \tiny 0.51	 &\tiny 90.4  & \tiny1.9 & \tiny\cite{ref47} \\  
			\tiny	4	& \tiny 0.01	 &\tiny 69  & \tiny 12 & \tiny \cite{ref43}   & \tiny	27	& \tiny 0.57	 &\tiny 96.8  & \tiny3.4& \tiny \cite{ref55}  \\	
			\tiny	5	& \tiny 0.12	 &\tiny 68.6  & \tiny26.2 & \tiny \cite{ref52}  & \tiny	28	& \tiny 0.593	 &\tiny 104 & \tiny 13 & \tiny \cite{ref45}  \\ 
			%
			\tiny	6	& \tiny 0.17	 &\tiny 83  & \tiny 8 & \tiny \cite{ref43}   & \tiny	29	& \tiny 0.60	 &\tiny 87.9  & \tiny6.1 & \tiny\cite{ref49} \\ 
			%
			\tiny	7	& \tiny 0.179	 &\tiny 75  & \tiny 4  & \tiny \cite{ref45}  & \tiny	30	& \tiny 0.61	 &\tiny 97.3  & \tiny2.1 & \tiny\cite{ref47} \\ 		
			\tiny	8	& \tiny 0.1993	 &\tiny 75  & \tiny 5  & \tiny\cite{ref45}   & \tiny	31	& \tiny 0.68	 &\tiny 92  & \tiny 8 & \tiny\cite{ref45}   \\ 
			\tiny	9	& \tiny 0.2	 &\tiny 72.9  & \tiny 29.6  & \tiny \cite{ref52} & 	\tiny	32	& \tiny 0.73	 &\tiny 97.3 & \tiny 7 & \tiny \cite{ref40}   \\ 
			\tiny	10	& \tiny 0.24	 &\tiny 79.7  & \tiny 2.7 & \tiny \cite{ref54} & \tiny	33	& \tiny 0.781	 &\tiny 105 & \tiny 12 & \tiny \cite{ref45}   \\		
			\tiny	11	& \tiny 0.27	 &\tiny 77  & \tiny 14 & \tiny \cite{ref43}   & \tiny	34	& \tiny 0.875	 &\tiny 125  & \tiny 17 & \tiny \cite{ref45}  \\ 
			\tiny	12	& \tiny 0.28	 &\tiny 88.8  & \tiny 36.6 & \tiny \cite{ref52}  & 	\tiny	35	& \tiny 0.88	 &\tiny 90  & \tiny 40 & \tiny  \cite{ref43} \\
			%
			\tiny	13	& \tiny 0.35	 &\tiny 82.7 & \tiny 8.4 & \tiny \cite{ref46}    & \tiny	36	& \tiny 0.9	 &\tiny 117  & \tiny 23 & \tiny  \cite{ref43} \\ 
			\tiny	14	& \tiny 0.352	 &\tiny 83 & \tiny 14 & \tiny \cite{ref45}    & \tiny	37	& \tiny 1.037	 &\tiny 154  & \tiny 20 & \tiny \cite{ref45} \\ 
			\tiny	15	& \tiny 0.38	 &\tiny 81.5  & \tiny 1.9 & \tiny \cite{ref47}  & 	\tiny	38 	& \tiny 1.3	 &\tiny 168 & \tiny 17 & \tiny \cite{ref43} \\ 
			%
			\tiny	16	& \tiny 0.3802	 &\tiny 83 & \tiny 13.5 & \tiny\cite{ref48}  & 	\tiny	39	& \tiny 1.363	 &\tiny 160  & \tiny 33.6 & \tiny \cite{ref56} \\ 
			%
			\tiny	17	& \tiny 0.4	 &\tiny 95  & \tiny 17 & \tiny \cite{ref53}   & \tiny	40	& \tiny 1.43	 &\tiny 177  & \tiny 18 & \tiny  \cite{ref43} \\ 
			%
			\tiny	18	& \tiny 0.4004	 &\tiny 77 & \tiny10.2 & \tiny\cite{ref48}   & 	\tiny	41	& \tiny 1.53	 &\tiny 140  & \tiny  14 & \tiny \cite{ref43} \\ 
			\tiny	19	& \tiny 0.4247	 &\tiny 87.1  & \tiny 11.2 & \tiny\cite{ref48}  & 	\tiny	42	& \tiny 1.75	 &\tiny 202  & \tiny 40 & \tiny \cite{ref43}\\ 
			\tiny	20	& \tiny 0.43	 &\tiny 86.5  & \tiny 3.7 & \tiny\cite{ref54}  & \tiny	43	& \tiny 1.965	 &\tiny 186.5  & \tiny 50.4 & \tiny \cite{ref56}  \\ 
			\tiny	21	& \tiny 0.44	 &\tiny 82.6  & \tiny 7.8 & \tiny\cite{ref49}  & 	\tiny	44	& \tiny 2.3	 &\tiny 224 & \tiny 8 & \tiny\cite{ref57}  \\ 
			\tiny	22	& \tiny 0.44497	 &\tiny 92.8  & \tiny 12.9 & \tiny\cite{ref48}  & 	\tiny	45	& \tiny 2.34	 &\tiny 222 & \tiny 7 & \tiny\cite{ref58} \\
			%
			\tiny	23	& \tiny 0.47	 &\tiny 89 & \tiny49.6  & \tiny\cite{ref50}    & 	\tiny	46	& \tiny 2.36	 &\tiny 226 & \tiny 8 & \tiny\cite{ref59} \\
	    	\hline	
		 \end{tabular}
	\end{center}
\end{table}

\subsection{Pantheon data }

In this part, we fit the  Pantheon data (the latest compilation of SNIa with 40 binned in the redshift range $0.014 \leq z \leq 1.62)$ to get the best fit values of model parameters $n$ and $H_0$, by minimizing 
$\chi^{2} (\mu_{0}, n, H_{0})$ statistic: 

\begin{equation}\label{28}
\chi^{2} (\mu_{0}, n)=\sum _{i=1} \frac{[\text{$\mu $th}(\text{z}(\mu_{0},n,z_{i}))-\text{$\mu $obs}(z_{i})]^2}{\text{$\sigma $}(i)^2}
\end{equation}
``$\mu_{th}$ and $\mu_{obs} $" are represents as the theoretical and observed distance modulus. $\sigma(i)$ denotes  the standard error of 
the observed values.\\
The distance modulus $\mu(z)$ is read by
\begin{equation}\label{29}
\mu=m-M=\mu_{0}+5 \log_{10} D_{L} (z),
\end{equation}
where $m$ and $M$ are ``apparent magnitude and absolute magnitude" of any distant luminous object respectively. The  parameter $\mu_{0}$ and 
the luminosity distance $D_{L}(z)$ are defined as
\begin{equation}\label{30}
\mu_{0}=5 \log_{10} \left[\frac{H^{-1}_{0}}{Mpc}\right]+25,
\end{equation}
and 
\begin{equation}\label{31}
D_{L}(z)=(1+z) \int_{0}^{z} \frac{dz}{H(z)}.
\end{equation}
The absolute magnitude $M$ and apparent magnitude $m$ are read as
\begin{equation}\label{32}
M=16.08 -25+5\log_{10} \left(\frac{H_{0}}{0.026}\right),
\end{equation}
\begin{equation}\label{33}
m=16.08+5\log_{10} \left(\frac{(1+z)}{0.026} \int_{0}^{z} \frac{dz}{H(z)}\right).
\end{equation}

The best fit contour plots for the $H(z)$ data set  and Pantheon data set and their combined data set are shown in this figures $1$a, $1$b $\&$ $1$c.


\begin{figure}[H]
	\centering
	\begin{subfigure}[b]{0.3\textwidth}
		\centering
		\includegraphics[width=\textwidth]{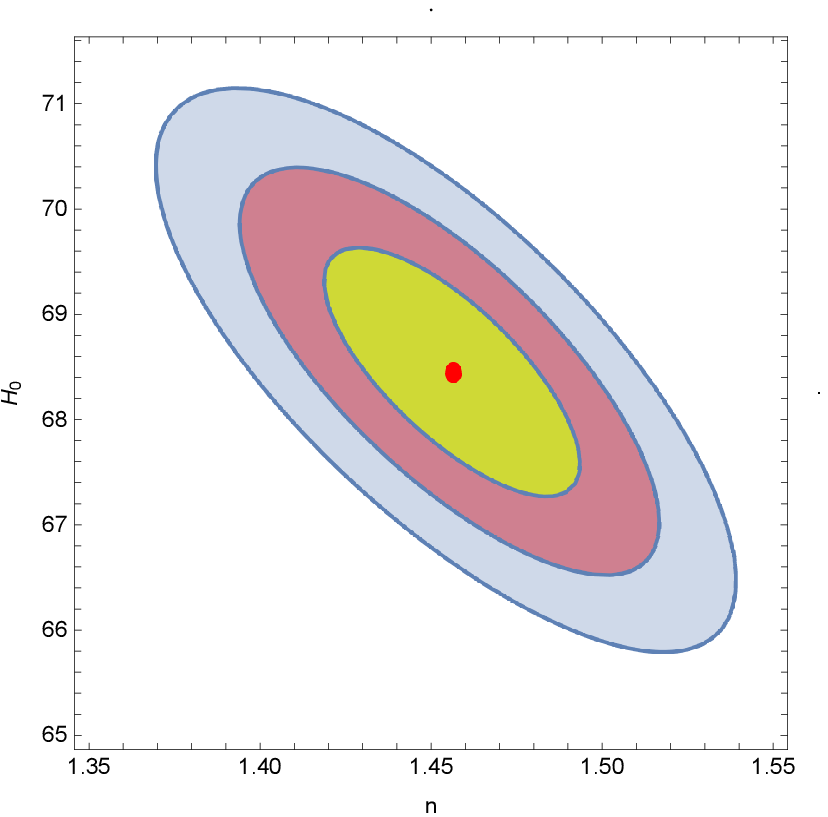}
		\caption{Contour plot for OHD}
		\end{subfigure}
	\hfill
	\begin{subfigure}[b]{0.3\textwidth}
		\centering
		\includegraphics[width=\textwidth]{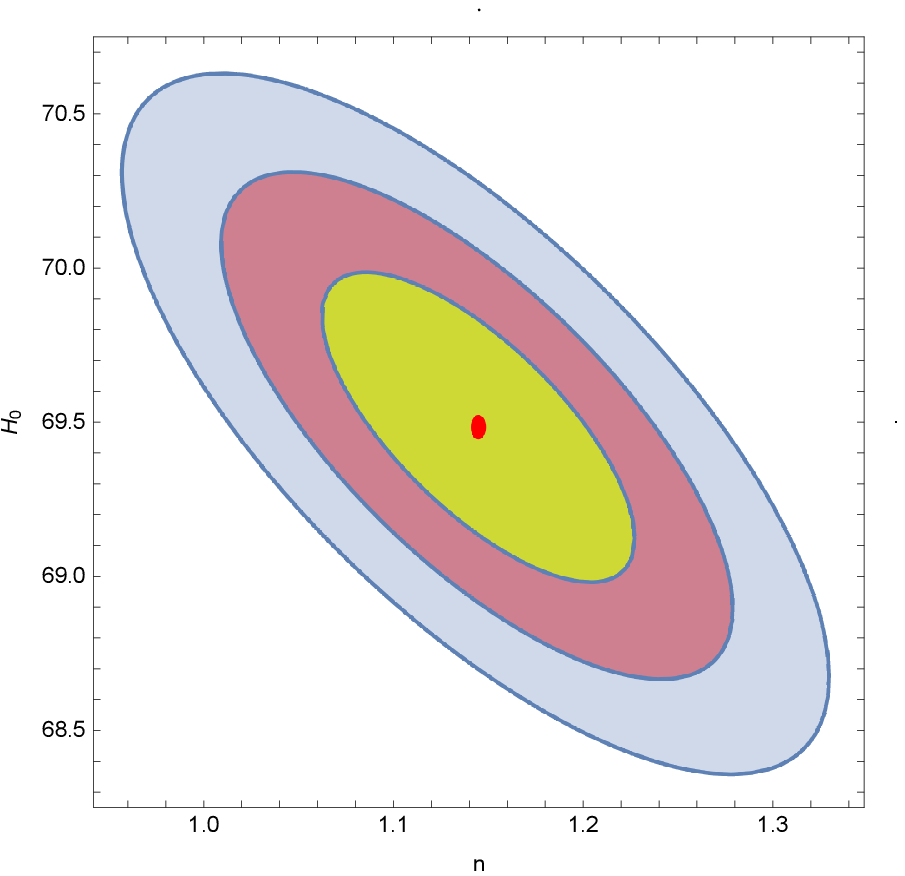}
		\caption{Contour plot for Pantheon data}
	\end{subfigure}
	\hfill
	\begin{subfigure}[b]{0.3\textwidth}
		\centering
		\includegraphics[width=\textwidth]{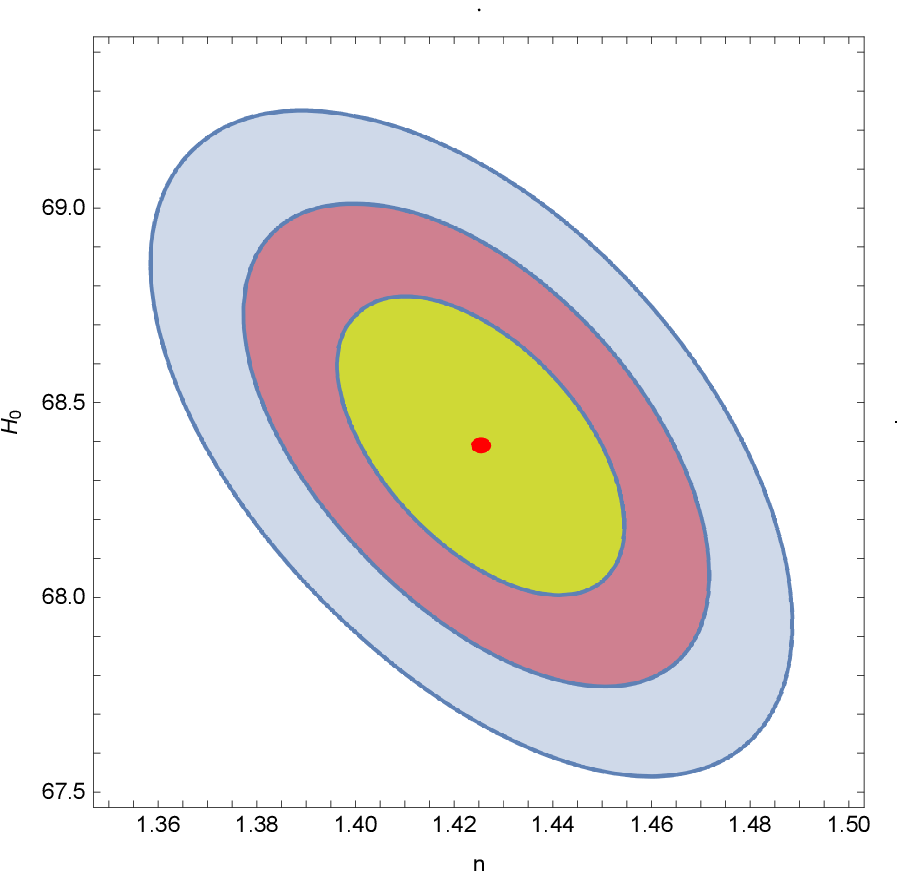}
		\caption{Contour plot for OHD+Pantheon}
	\end{subfigure}
	\caption{In this figure, the contour plots for likelihood values of model parameters $H _{0}$ $\&$ $n$  with samples of OHD and Pantheon data 
	and combined datasets are displayed at $1-\sigma$, $2-\sigma$, and $3-\sigma$ levels.}
	\label{fig1}
\end{figure}
\begin{figure}[H]
	\centering
	(a)\includegraphics[scale=0.6]{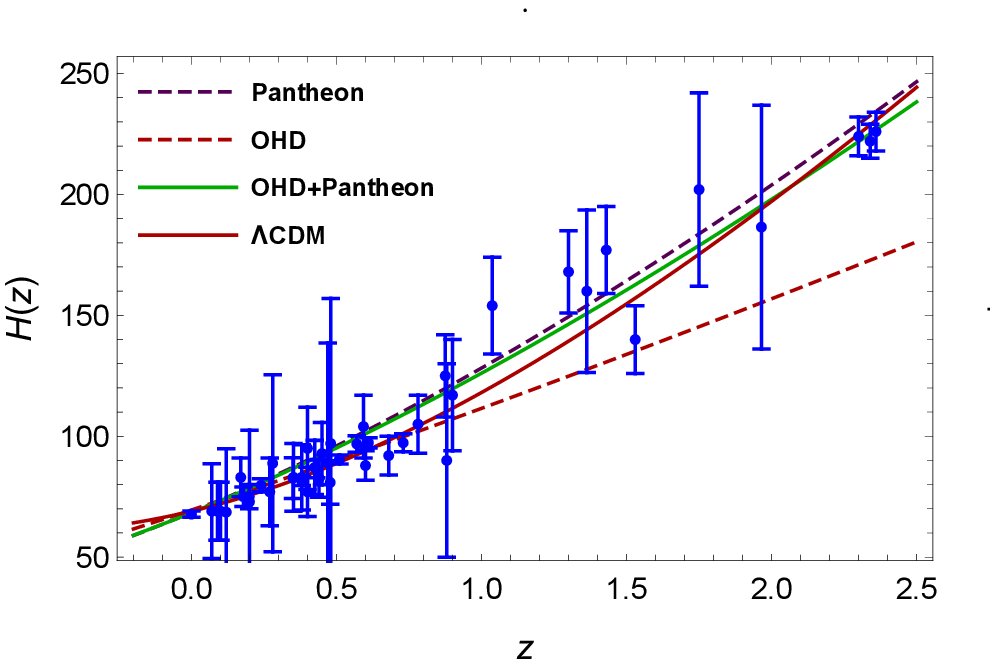}
	(b)\includegraphics[scale=0.6]{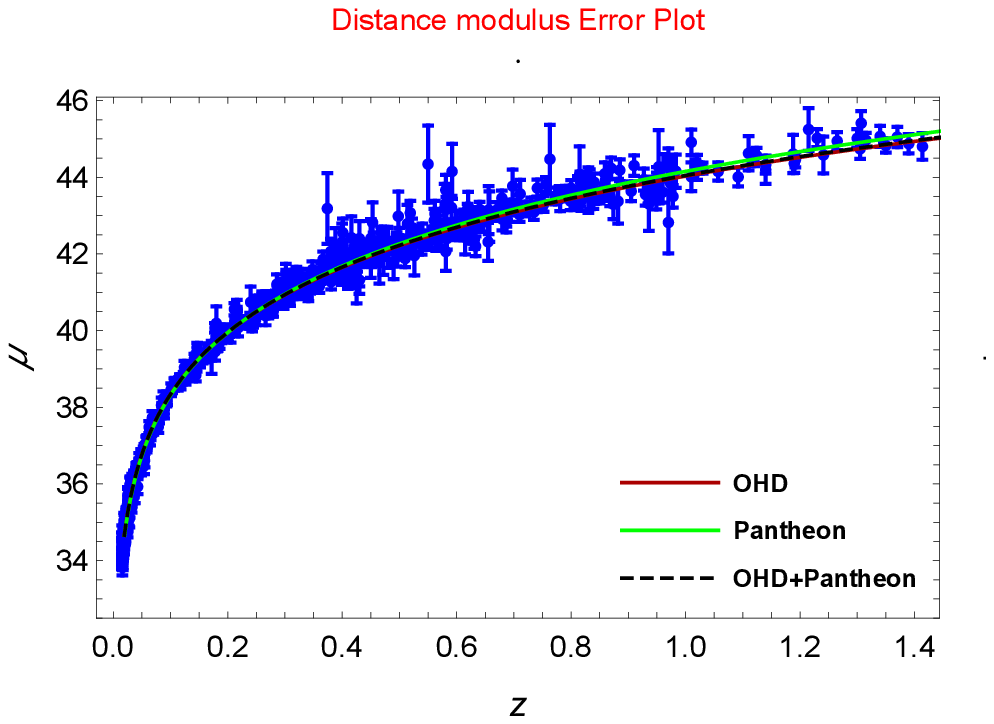}
	\caption{Fig.$2$a shows the comparison of theoretical model with error bar plots of the OHD data and the $ \Lambda$CDM model.
	Table 1 summarises the experimental data in the points with bars. When we perform joint analysis of two datasets, it is 
		evident that our generated model is best-fitted to data. The distance modulus error plot of the Pantheon data and the $\Lambda$CDM are shown in Figure $2$b, respectively.} 
		\label{fig2}	
\end{figure}
Using a differential age technique and galaxy clustering method, many cosmologists \cite{ref63,ref64} have computed the values of the Hubble 
constant at various red-shifts. They described several observed Hubble constant Hob values as well as corrections in the range $0 \leq z \leq 2.36$ 
\cite{ref40,ref65}. In our derived model the observed and theoretical values are found to agree quite well, indicating that our model is correct. 
The dots indicate the 46 observed Hubble constant (Hob) values with corrections, while the linear curves show the theoretical Hubble constant H(z) 
graphs with marginal corrections.

\section{Non-Interacting Model }

We suppose that DM and DE only interact gravitationally in the non-interacting model, therefore each source must satisfy the continuity equation independently. Thus, from Eq. (\ref{4}), we have 

\begin{equation}\label{34}
\dot{\rho_{m}}+3H \rho_{m}=0,
\end{equation}
\begin{equation}\label{35}
\dot{\rho_{de}}+3H \rho_{de}(1+\omega_{de})=0.
\end{equation}

The energy density of dark matter is calculated by using Eq. (\ref{34})
\begin{equation}\label{36}
 \rho_{m}=\rho_{0} a^{-3}=\rho_{0} \left(k_{1} e^{n \beta t} -1\right)^{\frac{-3}{n}}
\end{equation}

The energy density of dark energy is calculated by using Eqs. (\ref{23}) and (\ref{36}) in Eq. (\ref{20}).
\begin{eqnarray}\label{37}
&\rho_{de}=\phi _0 \left(k_{1} e^{\beta  n t}-1\right)^{\eta/n} \bigg[-\frac{c_1^2 \left(k_{1} e^{\beta  n t}-1\right)^{-6/n} 
\left(\left(k_{1}e^{\beta  n t}-1\right)^{1/n}\right)^{-2 \eta }}{\phi _0^2}-3 \alpha ^2 \left(k_{1} e^{\beta  n t}-1\right)^{-2/n}\nonumber\\
&-\frac{k^{2}_{1}\beta ^2 \left(\eta ^2 \omega -6 \eta -6\right) e^{2 \beta  n t}}{2 \left(k_{1} e^{\beta  n t}-1\right)^2}\bigg]-
\rho _{0} \left(k_{1} e^{\beta  n t}-1\right)^{-3/n}
\end{eqnarray}
Using the values of $\phi$ and $\rho_{de}$, the value of EoS parameter is obtained as 
\begin{eqnarray}\label{38}
&\omega_{de}=\frac{-\bigg[\frac{c_1^2 \left(k_{1} e^{\beta  n t}-1\right)^{-6/n} \left(\left(k_{1}e^{\beta  n t}-1\right)^{1/n}\right)^{-2 \eta }}
{\phi _0^2}-\alpha ^2 \left(k_{1} e^{\beta  n t}-1\right)^{-2/n}+\frac{k_{1}\beta ^2 e^{\beta  n t} 
\left(\left(\eta ^2 (\omega +2)+4 \eta +6\right) k_{1}e^{\beta  n t}-2 (\eta +2) n\right)}{2 \left(k_{1} e^{\beta  n t}-1\right)^2}\bigg]}
{ \bigg[-\frac{c_1^2 \left(k_{1} e^{\beta  n t}-1\right)^{-6/n} \left(\left(k_{1} e^{\beta  n t}-1\right)^{1/n}\right)^{-2 \eta }}{\phi _0^2}-
3 \alpha ^2 \left(k_{1}e^{\beta  n t}-1\right)^{-2/n}-\frac{k^{2}_{1} \beta ^2 \left(\eta ^2 \omega -6 \eta -6\right) e^{2 \beta  n t}}
{2 \left(k_{1}e^{\beta  n t}-1\right)^2}\bigg]-\frac{\rho _{0}}{\phi_{0}} \left(k_{1}e^{\beta  n t}-1\right)^{-(3+\eta)/n}}.
\end{eqnarray}


\section{Interacting Model }

We discuss the interacting cosmological models of the universe by considering the interaction between DE and DM components. As a result, the DM and DE 
the continuity equations are written as

\begin{equation}\label{39}
3H \rho_{m}+\dot{\rho_{m}}-Q = 0,
\end{equation}
\begin{equation}\label{40}
3H \rho_{de}(1+\omega_{de})+\dot{\rho_{de}}+Q = 0.
\end{equation}

The coupling between DM and DE is denoted by $Q$. We estimate the coupling between DE and DM as a function of  $\rho_{m}$ and 
$H$, $Q \propto \rho_{m} H$. As a result, for the interaction model, we use ``$Q = 3 b^2 \rho_{m} H$";  $b^{2}$ is a constant. \\
Now, the energy densities of DM and DE are determined as follows:
\begin{equation}\label{41}
\rho_{m}=k \left(k_{1}e^{\beta  n t}-1\right)^{\frac{3 \left(b^2-1\right)}{n}}.
\end{equation}
Here, $k$ is an integration constant.
\begin{eqnarray}\label{42}
&\rho_{de}=\phi _0 \left(k_{1} e^{\beta  n t}-1\right)^{\eta/n} \bigg[-\frac{c_1^2 \left(k_{1} e^{\beta  n t}-1\right)^{-6/n} 
	\left(\left(k_{1}e^{\beta  n t}-1\right)^{1/n}\right)^{-2 \eta }}{\phi _0^2}-3 \alpha ^2 \left(k_{1} e^{\beta  n t}-1\right)^{-2/n}\nonumber\\
&-\frac{k^{2}_{1}\beta ^2 \left(\eta ^2 \omega -6 \eta -6\right) e^{2 \beta  n t}}{2 \left(k_{1} e^{\beta  n t}-1\right)^2}\bigg]-
k \left(k_{1} e^{\beta  n t}-1\right)^{\frac{3 \left(b^2-1\right)}{n}}
\end{eqnarray}
For interacting model, the EoS parameter is obtained as:
\begin{eqnarray}\label{43}
&\omega_{de}=\frac{-\bigg[\frac{c_1^2 \left(k_{1} e^{\beta  n t}-1\right)^{-6/n} \left(\left(k_{1}e^{\beta  n t}-1\right)^{1/n}\right)^{-2 \eta }}
	{\phi _0^2}-\alpha ^2 \left(k_{1} e^{\beta  n t}-1\right)^{-2/n}+\frac{k_{1}\beta ^2 e^{\beta  n t} 
		\left(\left(\eta ^2 (\omega +2)+4 \eta +6\right) k_{1}e^{\beta  n t}-2 (\eta +2) n\right)}{2 \left(k_{1} e^{\beta  n t}-1\right)^2}\bigg]}
{ \bigg[-\frac{c_1^2 \left(k_{1} e^{\beta  n t}-1\right)^{-6/n} \left(\left(k_{1} e^{\beta  n t}-1\right)^{1/n}\right)^{-2 \eta }}{\phi _0^2}-
	3 \alpha ^2 \left(k_{1}e^{\beta  n t}-1\right)^{-2/n}-\frac{k^{2}_{1} \beta ^2 \left(\eta ^2 \omega -6 \eta -6\right) e^{2 \beta  n t}}
	{2 \left(k_{1}e^{\beta  n t}-1\right)^2}\bigg]-\frac{k}{\phi_{0}} \left(k_{1}e^{\beta  n t}-1\right)^{\frac{3 \left(b^2-1-\eta\right)}{n}}}.
\end{eqnarray}


\section{Physical properties and dynamical behaviour of model}
The evolution of density parameter for DM and DE are shown in Fig. 3

\begin{figure}[H]
	\centering
	(a)\includegraphics[scale=0.55]{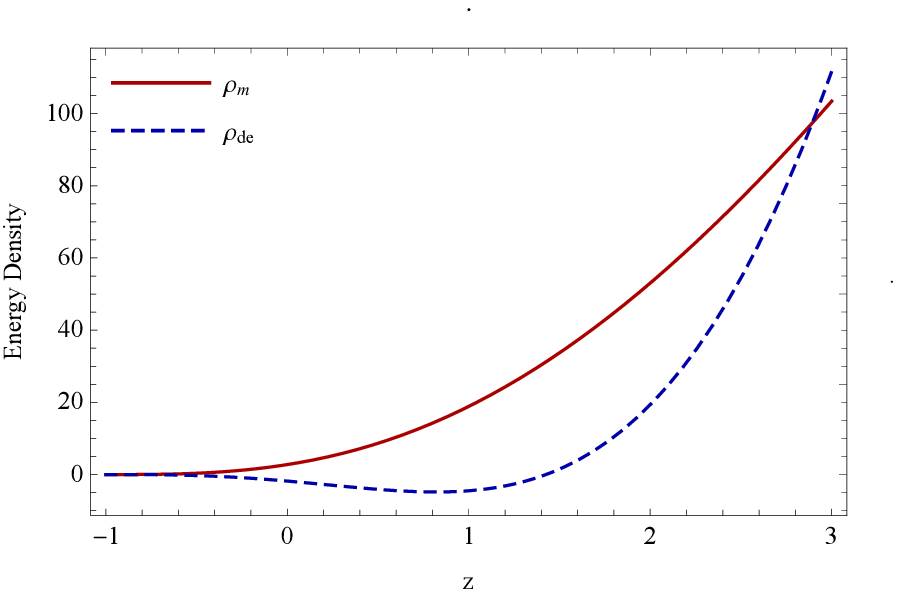}
	(b)\includegraphics[scale=0.55]{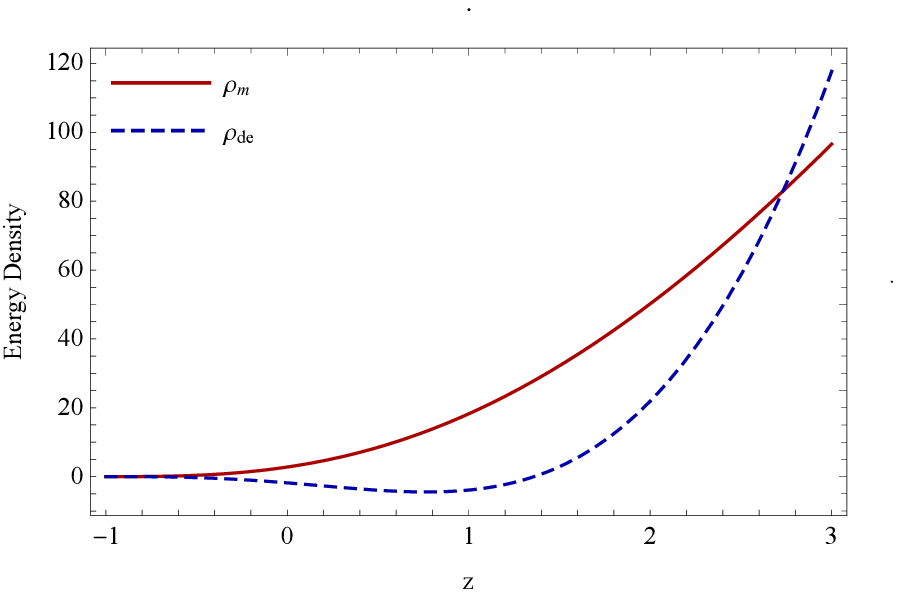}
	(c)\includegraphics[scale=0.55]{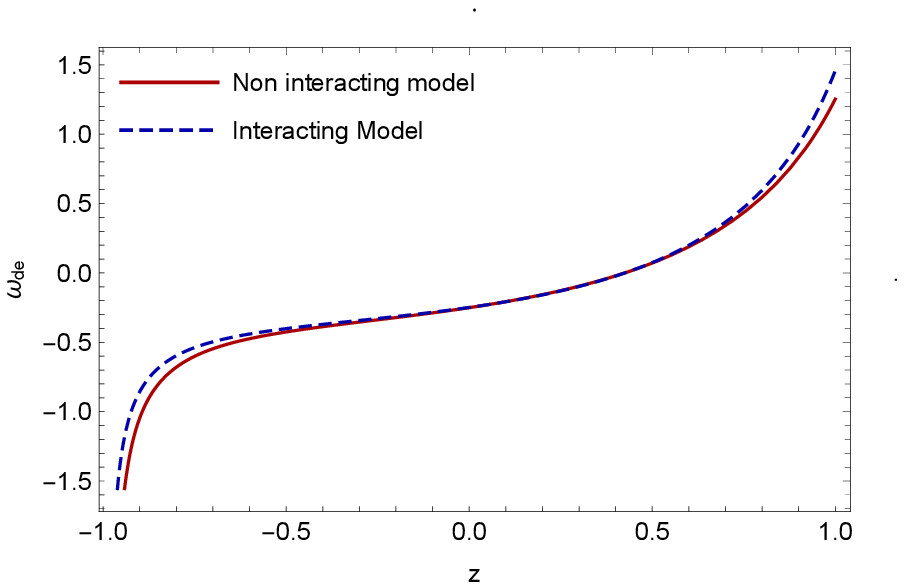}
	\caption{
	Plots of DM and DE energy densities versus redshift $z$ for non-interacting (left) and interacting (right) models $\&$ EoS parameter}
	\label{fig3}	
\end{figure}
Figure 3(a) shows the energy density for dark matter for the interacting and non-interacting cases with respect to redshift. 
Figure (3b) shows the variation of energy density for DE versus redshift. In both figures, we conclude that  as
$\rho\to0$ for (or $z\to -1$) which is consistent with a well-established scenario. From figure 3c, it is clear that the EoS of DE 
is varying in quintessence region and crossing the Phantom Divide line (PDL) $\omega_{de} =-1$ and finally enters in 
the phantom region in both interacting and non-interacting cases. The curve exhibits a negative tendency before evolving through several 
stages of acceleration, deceleration, and finally reaching to phantom phase. 

 The possibility of universe evolution is indicated by the shift from one phase to the next.


\subsection{Age of universe}
The age of the universe is obtained as
\begin{equation}\label{44}
dt=-\frac{dz}{(1+z) H(z)}\implies \int_{t}^{t_{0}} dt=-\int_{z}^{0} \frac{1}{(1+z) H(z)} dz .
\end{equation}
Using Eq. (\ref{26}), we get
\begin{equation}\label{45}
t_{0}-t=\int_0^z \frac{2}{H_{0} (z+1)\left[(z+1)^n+1\right]} \, dz,
\end{equation}
where $ t_{0} $ is the ``present age of universe" and it is given by
\begin{equation}\label{46}
t_{0}=\lim_{x \to \infty}\int_0^x \frac{2}{H_{0} (z+1)\left[(z+1)^n+1\right]} \, dz.
\end{equation}
Integrating Eq.(\ref{43}), we get
\begin{equation}\label{47}
H_{0} \ t_{0} = 0.9425
\end{equation}
In this study, we have calculated the numerical value of $ H_{0} $ as $ 0.07173 $ $ Gyr^{-1}$ $ \sim $ $70.29$ $ km s^{-1} Mpc^{-1} $. 
Therefore, the present age of the universe for the derived model is obtained as $ t_{0}=\frac{0.9425}{H_{0}}=13.14 $  Gyrs. 
Figure 7 shows the variation of ``$ H_{0} (t_{0}-t) $" with redshift $ z $. According to WMAP data \cite{ref34}, the empirical value of current age of the universe is $ t_{0}=13.73^{+.13}_{-.17} $.  It's worth noting that $t_0 = 13.81 \pm 0.038 $ Gyrs is the current age of the universe in Planck collaboration results \cite{ref65a}.  The universe's present has been calculated as $14.46\pm 0.8$ Gyrs  \cite{ref72}, $14.3 \pm 0.6$ Gyrs \cite{ref73} and $14.5 \pm 1.5$ Gyrs \cite{ref74} in various cosmological studies.
\begin{figure}[H]
	\centering
	\includegraphics[scale=0.9]{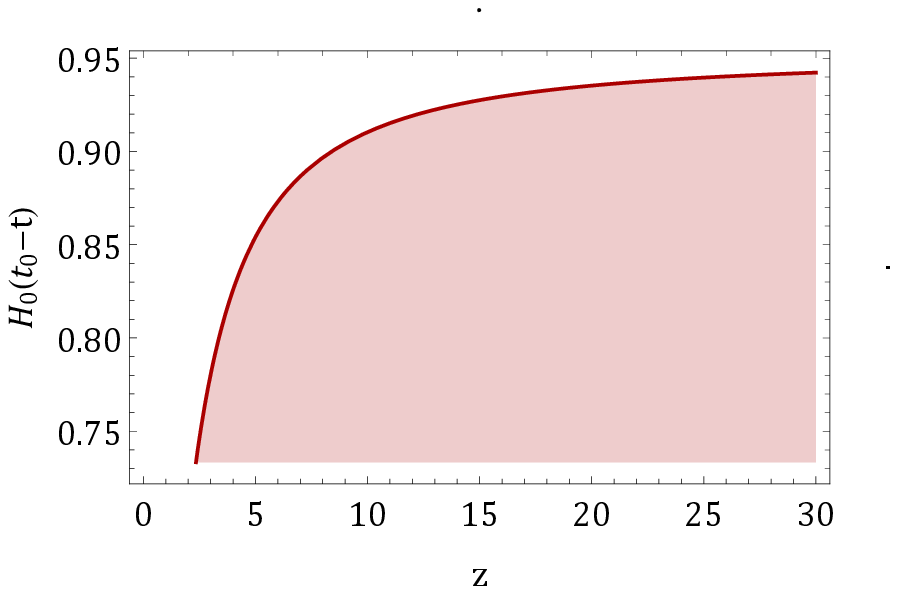}
	\caption{The plot of $H_{0} (t_{0}-t)$ versus $ z $ }\label{fig4}	
\end{figure}

\subsection{Deceleartion parameter}
Figure 5 depicts the dynamics of the decelerating parameter $(q)$ with respect to $z$. The models for OHD and (OHD+Pantheon) coincide with $\Lambda$CDM  with the signature flipping at $z_{t}=0.809$ while for Pantheon data model shows signature 
flipping at $z_{t}=4.44123$ due to the dominance of DE in the universe. As a result, the current universe evolves with a negative sign of $q$, 
causing  the accelerated expansion of the universe. Moreover the reconstruction of $q(z)$ is done by the joined $(SNIa + CC + H_{0})$, which have obtained the transition redshift $z_{t} = {0.69}^{+0.09}_{-0.06}, {0.65}^{+0.10}_{-0.07}$ and ${0.61}^{+0.12}_{-0.08}$ within  $(1 \sigma)$  \cite{ref65b}. which are seen
as well consistent with past outcomes \cite{ref65c,ref65d,ref65e,ref65f,ref65g} including the $\Lambda CDM$ expectation $z_{t} \approx 0.7$. $0.6 \leq z_{t} \leq 1.18$ ($2\sigma$, joint examination ) \cite{ref65h} is the other limit of transition redshift.\\

The recent 36 observational Hubble data (OHD) provides the redshift range $0.07 \leq z \leq 2.36$ \cite{ref65i}. The joint light curves (JLA) sample, comprised of 740 type-Ia supernovae (SN Ia) indicates the redshift range $0.01 \leq z \leq 1.30$. In this way, our developed model depicts a  transition from the early deceleration phase to the current speeding phase for OHD, Pantheon, (OHD+Pantheon) data. 
Furthermore, we find that the deceleration parameter will remain negative in the future, $z\to-1$, $q\to-1$.\\

The DP in terms of the redshift $z$ can be written as

\begin{equation}\label{48}
	q(z)=\frac{(n-1)(1+z)^{n}-1}{1+(1+z)^{n}}
\end{equation}

\begin{figure}[H]
	\centering
	\includegraphics[scale=0.9]{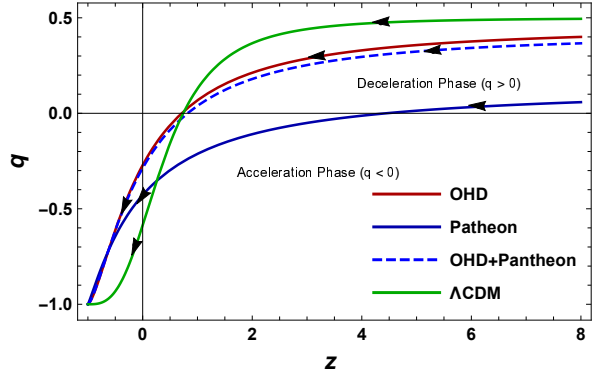}
	\caption{Plot of  deceleration parameter $ q $ versus $ z $. }\label{fig5}	
\end{figure}
\subsection{Luminosity Distance}

The redshift-luminosity distance connection  \cite{ref66} is a significant observational tool for studying the universe's evolution. 
The luminosity distance $(D _{L})$ is calculated in terms of the redshift, which occurs when light coming out of a  distant luminous body is 
redshifted due to the expansion of the cosmos. The luminosity distance is used to calculate a source's flux. It's written as $D_L = a_0 r(1 + z)$, 
where $r$ denotes the radial coordinate of the source. For the model, the radial coordinate $r$ is obtained as 
$r =\int_{0}^{dr} = \int_{0}^{t} \frac{c dt}{a(t)}=\frac{1}{a_{0} H_{0}} \int_{0}^{z} \frac{c dz}{h(z)}$, where $h(z)=\frac{H(z)}{H_0}$.\\

Therefore, we get the luminosity distance as
\begin{equation}\label{49}
\frac{H_{0} D_{L}}{c}=(1+z) \int_{0}^{0} \frac{dz}{h(z)}
\end{equation}

\subsection{Particle horizon}
The particle horizon is a measurement of the size of the observable cosmos \cite{ref67}.  The particle horizon is reads as
\begin{equation}\label{50}
R_P=\lim_{t_{P} \to 0} a_{0} \int_{t_{0}}^{t_{p}} \frac{dt}{a(t)}=\lim_{z \to \infty} \int_{0}^{z} \frac{dz}{H(z)}
\end{equation}
where $t_p$ denotes the "time in past" at which the light signal is sent from the source.
When we integrate Eq. (29) for a large value of red-shift, we get $R_P= 2.7845$ $H_{0}^{-1}$ as the particle horizon. Figure 6 depicts the 
dynamics of correct distance vs red-shift. We can see from Figure 7 that when  $z = 0$), $a_{0} H_{0} x$ is null, implying that when $z = 0$, 
the appropriate distance x becomes infinite. Thus we are at very large distance ( $\sim$ at infinite distance) from an event occurred in the 
beginning of the universe.

\begin{figure}[H]
	\centering
	\includegraphics[scale=0.9]{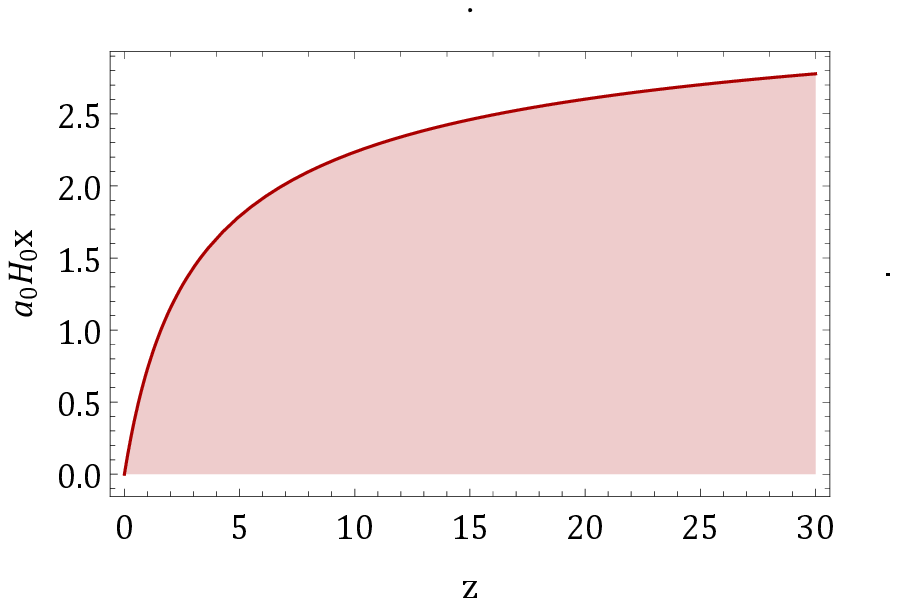}
	\caption{Distance of particle horizon versus $ z $ }\label{fig6}	
\end{figure}

\section{Statefinder diagnostics}

The statefinder pairs $\{r,s\}$ are the geometrical quantities that are directly obtained from the metric. This diagnostic is used to distinguish
different dark energy models and hence becomes an important tool in modern cosmology. The statefinder parameters $r$ and $s$ have been defined
as follows \cite{ref68,ref69,ref70}

\begin{equation}\label{51}
r=\frac{\dddot{a}}{a H^3},   \   s=\frac{r-1}{3 (q-\frac{1}{2})}
\end{equation}

\begin{equation}\label{52}
	r = \frac{n^2 \left(2 (z+1)^n+1\right) (z+1)^n}{\left((z+1)^n+1\right)^2}-\frac{3 n (z+1)^n}{(z+1)^n+1}+1
\end{equation}
\begin{equation}\label{53}
	s = \frac{2 n (z+1)^n \left(2 n (z+1)^n-3 \left((z+1)^n+1\right)+n\right)}{3 \left((z+1)^n+1\right) \left(2 n (z+1)^n-3 
	\left((z+1)^n+1\right)\right)}
\end{equation}

\begin{figure}[H]
	\centering
	(a)\includegraphics[scale=0.5]{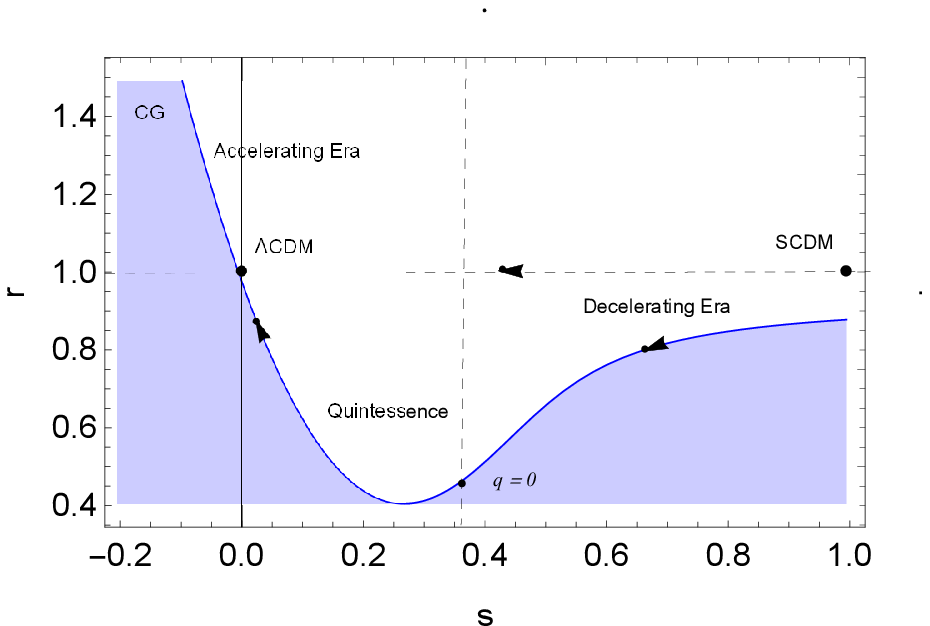}
	(b)\includegraphics[scale=0.5]{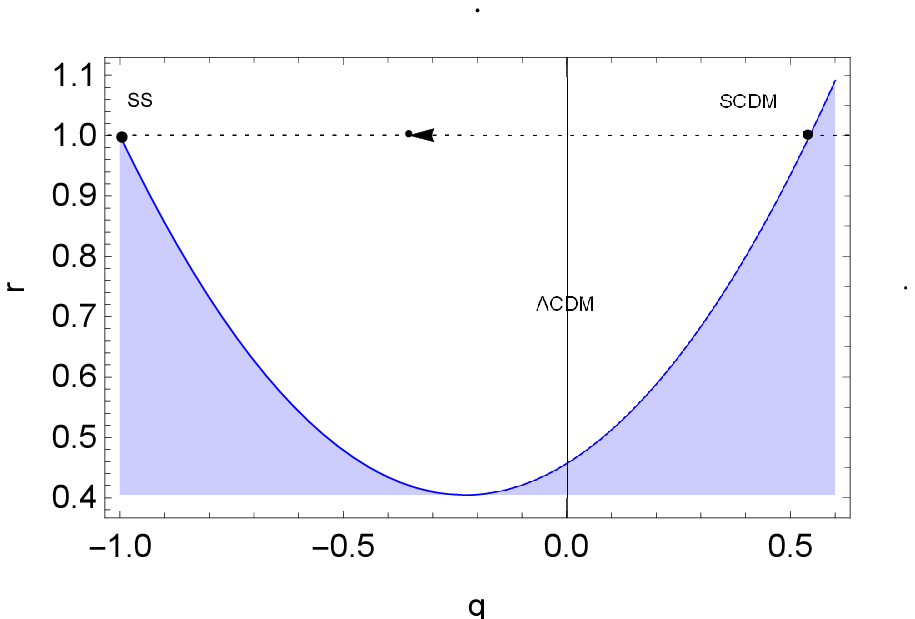}
	(c)\includegraphics[scale=0.5]{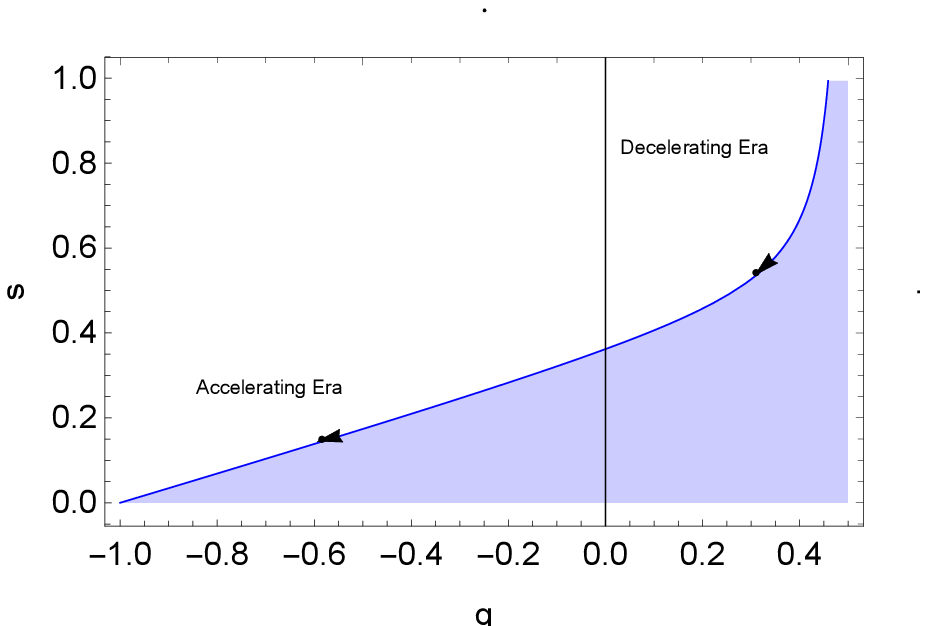}
	\caption{
	Plots of statefinders $(r~-s)$ $(r~-q)$ and $(s~-q)$ versus redshift $z$. }
	\label{fig7}	
\end{figure}

The statefinder diagnostic is a valuable tool in current cosmology used to differentiate various dark energy models \cite{ref68,ref69,ref70,ref71}. 
Different trajectories in the $(r-s)$ $(r- q)$ and $(s-q)$ planes express the time evolution for various dark energy models. The present model 
initially lies in quintessence region ($r<1, s> 0$), passes through the $\Lambda$CDM $(r, s) = (1, 0)$ and finally approaches to Chaplygin gas region
$r > 1$ and $s<0$ as clearly seen in Fig.7a.
The statefinder pair for $\Lambda$CDM and standard cold dark matter(SCDM) are respectively $(r, s) = (1, 0)$ and $(r, s) = (1, 1)$  depicts in 
the framwork of Bianchi-V background.
To get more information on the parametrization, we analyzed the temporal growth of our model by the $(r- q)$ plane.  
In Fig. 7(b) The solid line in the diagnostic plane's shows the evolution of the $\Lambda$CDM cosmological model. Initially our model 
start with the SCDM point  $(r, q) = (1, -0.5)$ passing the $\Lambda$CDM $(r, q) = (1, 0)$ point and finally reaches to SS point $(r, q) = (1,-1)$. 
The SCDM universe is clearly represented by the $q >0$ and $r > 1$ area of the profile. The $s-q$ trajectory begins with a decelerating period and 
progress to an accelerating era (see Fig.7c).


\begin{table}[H]
	\caption{\small Constrained values of the model parameters with minimum $\chi^{2}$ values }
	\begin{center}
		\begin{tabular}{|c|c|c|c|c|c|}
			\hline
			\tiny	Datasets &	\tiny  $n$ & \tiny $H_{0}$ & \tiny $q (z=0)$ & \tiny $z_{t}$ & \tiny $\chi^2$\\
			\hline
			\tiny	$H(z)$	& \tiny 1.457$\pm 0.042$	  &\tiny 68.46$\pm 1.97$ & \tiny -0.2715& \tiny 0.711645 & \tiny 31.8169 \\
			\hline
			\tiny	Pantheon	& \tiny 1.144$\pm 0.085$  &\tiny 69  $\pm 0.49$   & \tiny -0.428& \tiny 4.44123& \tiny 569.787 \\  
			\hline
			\tiny	$H(z)+$Pantheon	& \tiny 1.426$\pm 0.030$	 &\tiny 68.39 $\pm 0.40$   & \tiny -0.287 & \tiny 0.8192& \tiny 628.552  \\  
		    \hline		
			\end{tabular}
	\end{center}
\end{table}

\section{Conclusion}

In this paper, first, we observed the cosmological parameters for the observable universe in Bianchi type V space time in Branc Dicke theory. 
Second, we used statistical $\chi^2$ tests to limit the various model parameters of the universe in the resultant model. Table-2 summaries the 
major findings of the statistical analysis.

The main objective of this article is to use independent observables such as Pantheon dataset, OHD, and their joint combination (OHD+Pantheon) 
to constrain the free parameters of the theoretical models, which potentially increase the sensitivity of our estimations. Out of these 
measurements, we specifically discuss the $H_{0}$ determination. 
The main features of the derived model are discussed in the following manner:
\begin{itemize}
	\item 
	First, we have found a perfect solution to Einstein's field equations for both interacting and non-interacting cases in Bianchi type V space-time.
	\item 
	We have calculated best-fitted values of the model parameters from the marginal 1$\sigma$, 2$\sigma$, and 3$\sigma$ contour plots of OHD, Pantheon, 
	and (OHD+Pantheon) as shown in Fig.1. The constraints on $H_{0}$ and $n$ for the derived model by fitting the OHD points and the estimate 
	of Hubble's constant agree closely with Riess et al.\cite{ref1}.
	
	\item 
	In Fig. $2$a, we have shown the error bar plots for OHD data set, theoretical model plot for OHD data, and the $ \Lambda$CDM model. The points with bars 
	indicate the experimental data summarized in Table-1. The 46 points of the $H(z)$ and Union 2.1 compilation data are used to constrain 
	the model parameter $n$ and $H_0$. The derived model agrees well with $H(z)$ and Pantheon data and closely resembles the behavior of the $\Lambda$CDM. The distance modulus error plot of the Pantheon data alongwith the $\Lambda$CDM 
	are shown in Figure $2$b.
	
	\item  
	 Figures 3a and 3b show the graphical behavior of the energy densities for DM and DE. We noticed  that $\rho_{m}$ and $\rho_{d}$ of the model under discussion are positive decreases with time and vary individually with redshift in both interacting and non-interacting scenarios. Figure 3c shows the variation of the DE ($\omega_{de})$ equation of state parameter versus redshift for interacting and non-interacting models.
	
	It has been observed that $(\omega_{de})$ begins in the positive and ends in the negative. Note that $\omega_{de}>-1$ and $\omega_{de}<-1$ 
	respectively indicate quintessence and phantom regimes. While $\omega_{de}=-1 $ represents a universe dominated by cosmological constant. 
	This nature of EoS is ruled out by SN Ia findings. As a result, 
	the evolving range of $\omega_{de}$ of our derived model favors the phantom universe in the present epoch.
	
	\item
	The age of the universe in the derived model is $13.14$ Gyrs, as determined by OHD observations (see Fig. 4). The current age of the universe is believed to be $t_{0} = 13:65$ Gyrs. This age of the cosmos is very similar to the results found through Plank's observation\cite{ref65a} and several other observations\cite{ref72,ref73,ref74}.
	
	\item In our derived model, there is a smooth transition from decelerating to accelerating phase, which is consistent with the present scenario of the modern cosmology (see Fig. 5).
	We have plotted the deceleration parameter $q$ by getting its numerical solution, which exhibits a transition at $z_{t} = 0.711$, 
	from the early decelerated phase to a late time accelerated phase. This is in good agreement with the current cosmological observations.
	We have depicted that the deceleration parameter for OHD, (OHD+Pantheon) coincide with the $\Lambda$CDM model and have the same transition point(see Fig. 5).
	
	\item
	Particle horizon exists in the derived model, and its value differs from the $\Lambda$CDM universe model (see Fig.6). The age of the 
	universe in the developed model matches the empirical value obtained from WMAP measurements and Plank collaborations quite well.
	
	\item
	The variation of the trajectories $r-s$, $r-q$ and $s-q$ planes are depicted in Fig. 7. In the  diagram, the paths of the $r-s$ trajectories are depicted by arrows, which represent various dark energy models, and eventually approach to $\Lambda$CDM (see Fig. 7a). The evolution of the trajectories begins at SCDM in the $r-q$ plane diagram, and as time passes, the trajectories of the $r-q$ approach the steady-state model SS (see Fig. 7b). The $s-q$ trajectories begin with a decelerating period and progress to an accelerating era (see Fig. 7c).
\end{itemize}
	We noticed the range of the deceleration parameter from the equation of $q(z)$, which clearly demonstrates the signature flipping behavior. As a result, we find that the current OHD and Pantheon data give well-constrained $H _{0}$ values, and our model is consistent with recent observations. Furthermore, we describe the dynamics of the cosmos for both interacting and non-interacting scenarios.

\section*{Acknowledgments}
A. Pradhan is thankful to IUCAA, Pune, India for providing  support and facility under Visting Associateship prpgram.

\end{document}